\documentclass[aps,prd,notitlepage,nofootinbib,superscriptaddress,preprintnumbers,10pt]{revtex4-1}
\usepackage[latin9]{inputenc}
\usepackage{color}
\usepackage{bm}
\usepackage{amsmath}
\usepackage{amssymb}
\usepackage{graphicx}
\usepackage[unicode=true,bookmarks=false,breaklinks=false,pdfborder={0 0 1},backref=false,colorlinks=false]{hyperref}
\usepackage{breakurl}
\usepackage{epstopdf}
\usepackage{array}
\usepackage{natbib}

\makeatletter

\def\Mpl{M_{\rm P}}
\makeatother

\begin{document}

\preprint{YITP-18-45, IPMU18-0093}

\title{Phenomenology of minimal theory of quasidilaton massive gravity}

\author{Antonio De Felice}
\affiliation{Center for Gravitational Physics, Yukawa Institute for Theoretical Physics, Kyoto University, 606-8502, Kyoto, Japan}
\author{Shinji Mukohyama}
\affiliation{Center for Gravitational Physics, Yukawa Institute for Theoretical Physics, Kyoto University, 606-8502, Kyoto, Japan}
\affiliation{Kavli Institute for the Physics and Mathematics of the Universe (WPI), The University of Tokyo Institutes for Advanced Study, The University of Tokyo, Kashiwa, Chiba 277-8583, Japan}
\author{Michele Oliosi}
\affiliation{Center for Gravitational Physics, Yukawa Institute for Theoretical Physics, Kyoto University, 606-8502, Kyoto, Japan}

\date{\today}

\begin{abstract}
 The minimal theory of quasidilaton massive gravity with or without a Horndeski-type kinetic term for the quasidilaton field propagates only three physical modes: the two massive tensor polarizations and one scalar mode. This reduced number of degrees of freedom is realized by a Lorentz symmetry violation at cosmological scales and the presence of appropriate constraints that remove unwanted modes. Vacuum cosmological solutions have been considered in a previous work, and it has been shown that the late-time de Sitter attractor is stable under inhomogeneous perturbations. In this work, we explore the stability of cosmological solutions in the presence of matter fields. Assuming for simplicity that the quasidilaton scalar is on an attractor at the level of the background, we derive stability conditions in the subhorizon limit, and find the scalar sound speeds, as well as the modification with respect to general relativity to the gravitational potential in the quasistatic approximation. We also find that the speed limit of gravitational waves coincides with the speed of light for any homogeneous and isotropic cosmological background, on or away from the attractor. 
\end{abstract}

\maketitle

\section{Introduction}
One of the most important questions in modern cosmology remains the elucidation of the origin of the current accelerated expansion of the Universe \cite{acceUniv}. One of the studied avenues to address this so-called dark energy puzzle lies in large-distance, or infra-red (IR), modifications to general relativity (GR). For example, gravity could be weaker at distances larger than the separation between galaxy clusters so that a large cosmological constant would not gravitate as strongly as in GR and would lead to a modest acceleration as observed. Alternatively, in the absence of a cosmological constant an IR modified behavior of gravity could dynamically source the acceleration of the expansion. 

Among the IR modifications to Einstein gravity, adding a mass to the gravitational field is often presented as one of the simplest options, at least conceptually. In practice, it has however been difficult to construct a theory of massive gravity without compromising basic assumptions. In particular, the de Rham Gabadadze Tolley (dRGT) massive gravity \cite{dRGT}, unique healthy non-linear theory of pure massive gravity in four dimensions, does not accommodate Friedmann Lema\^{i}tre Robertson Walker (FLRW) solutions, meaning that either isotropy or homogeneity at large scales need to be considered as approximate \cite{nogoFLRWmassivegravity}. Although this does not compromise the validity of the theory as long as the breaking of the approximate symmetry occurs at large enough distances \cite{inhomanisomassivegravity}, it effectively renders the analysis more tenuous. In order to evade the practical issues, several extensions of dRGT massive gravity have thus since then been proposed, for instance: massive bigravity \cite{bigravity}, mass-varying massive gravity \cite{massvarying}, modified matter couplings \cite{MatterCouplingConstrainedVielbein}, scalar field extensions \cite{quasidilaton,extendedquasidilaton,generalized quasidilaton,newquasidilaton,chameleonbigravity}, and Lorentz symmetry violating extensions \cite{mtmg,mtmgph,mtmgphph,mmgt}. This has proven to be a fruitful path: several of these extensions are healthy and lead to interesting, non-$\Lambda$ cold dark matter (non-$\Lambda$CDM) phenomenology.

The minimal theory of quasidilaton massive gravity \cite{mqd,mqdh3}, by means of a weak violation of Lorentz invariance and the addition of a scalar field called the quasidilaton, has been shown i) to provide a stable cosmological vacuum de Sitter solution corresponding to a late-time de Sitter attractor, and ii) to accommodate the Vainshtein mechanism \cite{Vainshtein} to screen fifth force effects via a Horndeski-type kinetic term \cite{Horndeski} for the quasidilaton field. On a similar basis as the minimal theory of massive gravity (MTMG) \cite{mtmg}, the theory possesses constraints that remove the graviton scalar and vector modes. It thus propagates only three degrees of freedom, two tensors as in GR and the quasidilaton scalar, and can accommodate the Vainshtein screening for the quasidilaton. Therefore its phenomenology remains close to general relativity while giving a small mass (of cosmological value) to gravitational waves. 

In this work, we study both background cosmology and linear perturbations in the presence of matter fields, thus extending the work done in \cite{mqdh3}. We show that the dynamics of the quasidilaton scalar allow for an attractor that generalizes the attractor present in vacuum to the system with matter, analyze inhomogeneous perturbations around this attractor solution in the presence of matter, derive the stability conditions and find the expression for the effective gravitational constant in the quasistatic approximation.

The paper is organized as follows: in section \ref{sec:action} we present the complete action for the minimal theory of quasidilaton massive gravity, in the unitary gauge. In section \ref{sec:background} we present the cosmological background equations of motion and discuss the existence of the attractor solution. In section \ref{sec:perturbations} we study the behavior of linear perturbations on the attractor. Finally we summarize and discuss our results in section \ref{sec:discussion}.

\section{Action in the unitary gauge}\label{sec:action}

In order to construct a mass term for the physical metric field $g_{\mu\nu}$, one needs to introduce an additional, fiducial, metric $f_{\mu\nu}$. This non-dynamical metric is taken as part of the definition of the theory. In the minimal theory of quasidilaton massive gravity, thanks to the quasidilation symmetry of the action (\textit{c.f.}\ (\ref{eq:quasidilatation})), one may choose a Minkowski fiducial metric and retain stable cosmological solutions \cite{mqd,mqdh3}. For the remainder of the paper, we will thus keep $f_{\mu\nu} = \eta_{\mu\nu}$.

The graviton mass term, formed by contractions between the fiducial metric and the physical metric, will break the general covariance of the theory. While one may recover covariance by introducing four St\"{u}ckelberg scalar fields, in what follows we remain in the ``unitary gauge'': the gauge in which the St\"{u}ckelberg fields do not appear. Note that once this gauge is chosen, a change of coordinates will correspond, in general, to a different physical solution.

Out of the two metrics, one can read off the ADM lapse functions, shift vectors, three dimensional metrics, and their inverses as
\begin{gather}
N = \frac{1}{\sqrt{-g^{00}}}\,,\quad N^i = \frac{1}{N^2}g^{i0}\,,\quad\gamma^{ij} = g^{ij} + \frac{N^iN^j}{N^2}\,,\quad \gamma_{ik}\gamma^{kj} = \delta^j_i\,,\\
M = 1\,,\quad M^i = M_i=  0\,,\quad\tilde{\gamma}^{ij} = \tilde{\gamma}_{ij} = \delta_{ij}\,,
\end{gather}
where ($N$, $N^i$, $\gamma^{ij}$, $\gamma_{ij}$) correspond to the physical sector, ($M$, $M^i$, $\tilde{\gamma}^{ij}$, $\tilde{\gamma}_{ij}$) correspond to the fiducial sector, and $i,\,j,\,k\in\{1,2,3\}$. Using both three-dimensional metrics, one may further define $\mathfrak{K}^i{}_j$ and its inverse $\mathcal{K}^i{}_j$ such that
\begin{equation}
\mathfrak{K}^i{}_k\mathfrak{K}^k{}_j = \gamma^{ik}\tilde{\gamma}_{kj}\,,\qquad \mathfrak{K}^i{}_k\mathcal{K}^k{}_j = \mathcal{K}^i{}_k\mathfrak{K}^k{}_j = \delta^i_j\,.
\end{equation}

Building upon \cite{mqdh3}\footnote{We refer the curious reader to both \cite{mqd,mqdh3} for more details on the construction of the Lagrangian, and in particular the introduction of constraints and their Lagrange multipliers.}, the action for the minimal quasidilaton, in the presence of matter, may be written using the objects defined above as
\begin{equation}
S = \tilde{S}_{\textrm{EH}} + \tilde{S}_\sigma+ S_\textrm{pot} + S_\textrm{C} + S_\textrm{m}\,,
\end{equation}
where $\tilde{S}_{\textrm{EH}}$ is a part that includes the Einstein-Hilbert action for the physical metric $g_{\mu\nu}$ (without cosmological constant, since this one is included in the graviton mass term), $\tilde{S}_\sigma$ is a part that includes the kinetic term of the quasidilaton scalar field $\sigma$, $S_\textrm{pot}$ is the precursor graviton mass term, $S_\textrm{C}$ is a part that includes the additional constraints which remove the scalar graviton mode, and $S_\textrm{m}$ is the matter action. 

For compactness, both kinetic terms for the graviton and the quasidilaton may be supplemented by a contribution from the quasidilaton cubic Horndeski sector, whose tertiary constraint mixes with the supplementary constraints used to minimize the number of physical degrees of freedom in the theory \cite{mqdh3}. We thus rewrite the action in a way that includes these contributions into the kinetic terms, i.e.\ the Einstein-Hilbert action as
\begin{equation}
\tilde{S}_\textrm{EH} = \int d^4x N \sqrt{\gamma} \left(R^{(3)} + \tilde{K}_{ij}\tilde{K}^{ij}-\tilde{K}^2\right)\,, \label{eqn:SEH}
\end{equation}
and the quasidilaton kinetic term as
\begin{equation}
\tilde{S}_\sigma = \int d^4x \sqrt{-g}\,\left[P(X)-G(X)\Box\sigma+ \lambda_\chi\left(\tilde{\mathfrak{X}}_\sigma-X\right)\right]\,,\label{eqn:Ssigma}
\end{equation}
where $R^{(3)}$ is the three-dimensional Ricci curvature, $\lambda_\chi$ is a Lagrange multiplier, 
\begin{eqnarray}
\tilde{K}_{ij} & = & K_{ij} + \frac{1}{M_\textrm{P}^2}\frac{\lambda_T}{N}G_{,X}\left(X\gamma_{ij} + \sigma_{;i}\sigma_{;j}\right)\,,\nonumber\\
 \tilde{\mathfrak{X}}_\sigma& = &\frac{1}{2}\left[\widetilde{\partial_\perp\sigma}^2-\gamma^{ij}\partial_i\sigma\partial_j\sigma\right]\,,\quad
\widetilde{\partial_\perp\sigma} = \partial_\perp\sigma + \frac{\lambda_T}{N} = \frac{1}{N}\left(\dot{\sigma} - N^i\sigma_{;i}\right) + \frac{\lambda_T}{N}\,,
\end{eqnarray}
and $\lambda_T$ is the Lagrange multiplier for a tertiary constraint arising from the Horndeski sector. In the expressions (\ref{eqn:SEH})-(\ref{eqn:Ssigma}), to simplify the form of the total action $S$ we have replaced the usual extrinsic curvature $K_{ij}$ (and its trace $K = \gamma^{ij}K_{ij}$) by $\tilde{K}_{ij}$ (and its trace $\tilde{K} = \gamma^{ij}\tilde{K}_{ij}$), and replaced the usual canonical kinetic term for the quasidilaton scalar field $\mathfrak{X}_\sigma$ by $\tilde{\mathfrak{X}}_\sigma$.

From here on, we use a semicolon to indicate covariant derivatives w.r.t.\ $\gamma_{ij}$. The traces of $\mathcal{K}^i{}_j$ and $\mathfrak{K}^i{}_j$ are denoted as $\mathcal{K} \equiv \mathcal{K}^i{}_i$ and $\mathfrak{K} \equiv \mathfrak{K}^i{}_i$. Using these, the precursor graviton mass term is defined as
\begin{eqnarray}
S_{\rm pot} & = & \frac{M_{\mathrm{P}}^{2}}{2}\int d^4x\sum_{i=0}^{4}\mathcal{L}_{i}\,,\label{eq:lagrangian_gravitonpotential}\\
\mathcal{L}_{0} & = & -m^{2}c_{0}e^{(4+\alpha)\sigma/M_{\mathrm{P}}}\,\sqrt{\tilde{\gamma}}\,M\,,\label{eqn:def-S0}\\
\mathcal{L}_{1} & = & -m^{2}c_{1}e^{3\sigma/M_{\mathrm{P}}}\,\sqrt{\tilde{\gamma}}\,(N+Me^{\alpha\sigma/M_{\mathrm{P}}}\mathcal{K})\,,\\
\mathcal{L}_{2} & = & -m^{2}c_{2}e^{2\sigma/M_{\mathrm{P}}}\,\sqrt{\tilde{\gamma}}\,\left[N\mathcal{K}+\frac{1}{2}Me^{\alpha\sigma/M_{\mathrm{P}}}(\mathcal{K}^{2}-\mathcal{K}^{i}{}_{j}\mathcal{K}^{j}{}_{i})\right]\,,\\
\mathcal{L}_{3} & = & -m^{2}c_{3}e^{\sigma/M_{\mathrm{P}}}\sqrt{\gamma}\,(N\,\mathfrak{K}+Me^{\alpha\sigma/M_{\mathrm{P}}})\,,\\
\mathcal{L}_{4} & = & -m^{2}c_{4}\sqrt{\gamma}N\,. \label{eqn:def-S4}
\end{eqnarray}
The cosmological constant is included in this potential term and is proportional to the parameter $c_4$. 
The remainder constraints part of the action is further divided as
\begin{equation}
S_C = S_{\lambda^i} + \tilde{S}_{\lambda} + S_{\lambda^2}+ S_{\lambda_T}\,,
\end{equation}
where 
\begin{align}
S_{\lambda_i} & = \int d^4x\, \lambda^i \frac{m^2 M_\textrm{P}^2}{2}\left[\frac{1}{2}\sqrt{\gamma} \mathcal{D}_j\left(\Theta^{jk}\gamma_{ik}\right)-\mathcal{H}\partial_i\sigma\right],\\
\tilde{S}_\lambda & = - \int d^4x\,\lambda\frac{m^2 M_\textrm{P}^2}{4}\left(2\mathcal{H}\widetilde{\partial_\perp\sigma} + \sqrt{\gamma} \tilde{K}_{ij}\Theta^{ij}\right),\\
S_{\lambda^2} & = \int d^4x\,\lambda^2\frac{m^4 M_\textrm{P}^2}{64 N} \sqrt{\gamma}\left(2\Theta_{ij}\Theta^{ij} -\Theta^2\right),\\
S_{\lambda_T} & = \int d^4x\, N \sqrt{\gamma} \frac{\lambda_T}{N}\left[G_{,X}\widetilde{\partial_\perp\sigma}\,\sigma^{;i}{}_{;i} - G_{,X}\left(\widetilde{\partial_\perp\sigma}\right)^{;i}\sigma_{;i} -G_{,X}\partial_\perp X - P_{,X}\widetilde{\partial_\perp\sigma}\right],
\end{align} 
with $i\in \{1,2,3\}$. Here we have defined 
\begin{eqnarray}
\Theta^{ij} & = & e^{\alpha\sigma/M_{\mathrm{P}}}\left\{ \frac{\sqrt{\tilde{\gamma}}}{\sqrt{\gamma}}\left[\left(c_{1}e^{3\sigma/M_{\mathrm{P}}}+c_{2}e^{2\sigma/M_{\mathrm{P}}}\mathcal{K}\right)\left(\mathcal{K}^i{}_k\gamma^{kj}+\gamma^{ik}\mathcal{K}^j{}_k\right)-2c_{2}e^{2\sigma/M_{\mathrm{P}}}\tilde{\gamma}^{ij}\right]+2c_{3}e^{\sigma/M_{\mathrm{P}}}\gamma^{ij}\right\} \,,\\
\mathcal{H} & = & \frac{e^{\alpha\sigma/M_{\mathrm{P}}}}{M_{\mathrm{P}}}\left\{ \sqrt{\tilde{\gamma}}\left[(4+\alpha)c_{0}e^{4\sigma/M_{\mathrm{P}}}+(3+\alpha)c_{1}e^{3\sigma/M_{\mathrm{P}}}\mathcal{K}\right.\right.\nonumber \\
 &  & \left.\left.+\frac{2+\alpha}{2}c_{2}e^{2\sigma/M_{\mathrm{P}}}\,(\mathcal{K}^{2}-\mathcal{K}^{i}{}_{j}\mathcal{K}^{j}{}_{i})\right]+(1+\alpha)c_{3}\sqrt{\gamma}e^{\sigma/M_{\mathrm{P}}}\right\} \,.
\end{eqnarray}

Finally, for the matter action we assume that matter fields minimally couple to the $4$-dimensional physical metric, 
\begin{equation}
 g_{\mu\nu}dx^{\mu}dx^{\nu} = -N^2dt^2 + g_{ij}(dx^i+N^idt)(dx^j+N^jdt)\,, 
\end{equation}
constructed from ($N$, $N^i$, $g_{ij}$). Hereafter, for simplicity we consider a perfect fluid. In particular one may choose an action \`{a} la Schutz and Sorkin \cite{SorkinSchutz}, implemented as in \cite{mtmgph}
\begin{equation}
S_\textrm{m} = -\int d^4x \left(\sqrt{-g}\rho_m(n)+J^{\mu}\partial_{\mu}\varphi\right),
\end{equation}
where $n = \sqrt{\frac{J_\alpha J^\alpha}{-N^2\gamma}}$\, and $J^\mu$, $\phi$ are the (auxiliary) fields characterizing the fluid, or alternatively a k-essence scalar field
\begin{equation}
S_\textrm{m}' = \int d^4x \sqrt{-g}\, P_m (\mathfrak{X}_\textrm{m})\,,
\end{equation}
with $\mathfrak{X}_\textrm{m}=-\frac{1}{2}g^{\mu\nu}\nabla_\mu\phi_m\nabla_\nu\phi_m$ the canonical kinetic term for the matter scalar field $\phi_m$ and for which the density and the sound speed can be expressed as 
\begin{equation}
\rho_m = 2P_m'\mathfrak{X}_\textrm{m}-P_m\,,\qquad c_{s,m}^2 = \frac{P_m'}{2P_m''\mathfrak{X}_\textrm{m} + P_m'}\,.
\end{equation}
In the latter case, when discussing perturbations, one needs to change variables to density perturbations $\delta\rho$ before taking the dust limit ($P_m/\rho_m\to 0$, $c_{s,m}^2\to 0$), as needs to be done when computing the effective gravitational constant.

The total action is built to be invariant under the quasidilatation symmetry
\begin{equation}
\sigma \rightarrow \sigma + \sigma_0\,,\qquad M \rightarrow M e^{-(1+\alpha)\sigma_0/M_\mathrm{P}}\,,\qquad\tilde{\gamma}_{ij} \rightarrow \tilde{\gamma}_{ij} e^{-2\sigma_0/M_\mathrm{P}}\,,\label{eq:quasidilatation}
\end{equation}
with $\sigma_0$ a constant. It has been shown in \cite{mqd,mqdh3}, that this allows the theory to contain stable self-accelerating background cosmologies without the need for a time dependence in the fiducial metric.

\section{Cosmological background}\label{sec:background}

\subsection{Background equations}

Although this action is involved on arbitrary backgrounds, it has been shown in \cite{mqdh3} that on cosmological background, with line-element 
\begin{equation}
ds^2 = -N^2(t)dt^2 + a^2(t)\left(\delta_{ij}dx^idx^j\right),
\end{equation}
and where all fields are taken to be time-dependent only, the only solution is such that $\lambda= \lambda_T =\lambda^i = 0$, $i\in\{1,2,3\}$. For matter, we set $\rho_m=\rho_m^{(0)}(t)$ but for the ease of notation drop the superscript $(0)$. One can then obtain the background field equations from the mini-superspace action, which is greatly simplified compared to the general action. 

To treat cosmological backgrounds with more ease, we define 
\begin{equation}
\mathcal{X} = \frac{e^{\sigma/M_\textrm{P}}}{a}\,,\qquad r = \frac{a\, e^{\alpha\sigma/M_\textrm{P}}}{N}\,,\qquad \Sigma = \frac{\dot{\sigma}}{N}\,.
\end{equation}
The cosmological background equations are then written as
\begin{align}
E_1 &\equiv 3 M_\textrm{P}^2 H^2 - \rho_g - \rho_m = 0\,,\\
E_2 &\equiv M_\textrm{P}^2 \frac{2 \dot{H}}{N} + 3 M_\textrm{P}^2H^2 + P_g + P_m = 0\,,\\
E_\sigma &\equiv \frac{1}{2}\mathcal{X} M_\textrm{P}^2m^2[(\alpha + 1)rJ +rJ_{,\mathcal{X}}\mathcal{X} + \Gamma_{,\mathcal{X}}] + P_{,X}\Sigma \left(3 H + \frac{\dot{\Sigma}}{\Sigma N}\right)\nonumber\\
&+ \Sigma^3 \frac{\dot{\Sigma}}{\Sigma N}\left(P_{,XX}+G_{,XX}\Sigma H\right) + 3G_{,X}H\Sigma^2 \left(3H + \frac{\dot{H}}{H N} + 2\frac{\dot{\Sigma}}{\Sigma N}\right)= 0\,,\\
  E_X &\equiv G_{,X}\left(\frac{\dot{\Sigma}}{N}+3H\Sigma\right) +P_{,X} - \lambda_{\chi} = 0\,,\\
 E_{\lambda_T} & \equiv E_X \Sigma = 0\,,\\
 E_{\lambda_\chi} &\equiv X-\frac12\,\Sigma^2=0\,,\\
E_\lambda &\equiv \Gamma_{,\mathcal{X}}HM_\textrm{P} + \Sigma[(\alpha + 1)J + J_{,\mathcal{X}}\mathcal{X}] = 0\,,\label{eq:Elam}
\end{align}
whereas the equations for $\lambda^i$, $E_{\lambda^i}$, are trivially satisfied since they only include terms with spatial derivatives. We have defined the quantities
\begin{eqnarray}
\rho_g &\equiv&  2X P_{,X} - P + 6 H G_{,X}X\Sigma+\frac{M_\textrm{P}^2m^2}{2}\Gamma\,,\\
P_g &\equiv& P - 2G_{,X}X \frac{\dot{\Sigma}}{N} - \frac{M_\textrm{P}^2m^2}{2}\left[\Gamma - \frac{\Gamma_{,\mathcal{X}\mathcal{X}}}{3}(r-1)\right],
\end{eqnarray}
which can be interpreted as the contributions coming from the coupled sectors of the quasidilaton-field and the graviton-mass-term to the Friedmann equations. In these, for convenience we also have grouped 
\begin{gather}
\Gamma = \mathcal{X}^3c_1 + 3\mathcal{X}^2c_2+ 3\mathcal{X}c_3+ c_4\,,\qquad J = \mathcal{X}^3c_0 + 3\mathcal{X}^2c_1 + 3\mathcal{X}c_2 + c_3\,.
\end{gather}

\subsection{Cosmological attractor}\label{sec:attractor}
Just as in \cite{mqdh3}, from the equation for $\lambda$, one can infer the existence of a late-time de Sitter attractor. In fact, Eq.\ (\ref{eq:Elam}) can be rewritten as
\begin{equation}
  \frac{d}{dt}\,[a^{4+\alpha}\mathcal{X}^{1+\alpha}\,J(\mathcal{X})]=0\,.
  \end{equation}
  This result implies that
  \begin{equation}
    \mathcal{X}^{1+\alpha}\,J(\mathcal{X})=\frac{cst}{a^{4+\alpha}}\,,\label{eq:attractorspeed}
  \end{equation}

Notice that this result holds even in pure vacuum, i.e.\ this property is not changed by the presence of matter, since matter does not couple to $\lambda$. Notice that we do not consider the space-time to be de Sitter, rather, we will study the dynamics of the universe on an attractor solution. Such attractor solutions are defined by
\begin{equation}
\begin{cases}
\mathcal{X} = cst\,, & \textrm{for } \alpha = -4\,,\\
J = 0\,,\;\mathcal{X} = cst\,, \quad & \textrm{for } \alpha \neq -4\,.
\end{cases}\label{eq:attractor}
\end{equation}
On the attractor, one has thus further
\begin{equation}
\Sigma = H M_\mathrm{P}\,,\qquad X = \frac{M^2_\textrm{P}H^2}{2}\,.
\end{equation}
We will not study the fine-tuned case $\alpha = 4$ in this work. Instead, we will focus on the interval $\alpha>-4$, for which the system tends to the attractor solution $J=0$ --- having then neglected the strongly coupled solution $\mathcal{X}=0$. Since, in general, the solution differs from the attractor solution by an inverse power law, the rate of change of such a deviation from the attractor (in a Hubble time) is $-(4+\alpha)$, so that as long as $\alpha$ is not too close to -4, then the solution is quickly approximated by the attractor solution.

Finally, we notice here, that in practice, one can solve the three background equations of motion ($E_1$, $E_2$, and $E_\sigma$) for $P$, $r$, and $\dot{H}$. 

\section{Linear perturbations}\label{sec:perturbations}

We present here an analysis of the action quadratic in perturbations for the tensor, vector, and scalar sectors. The different helicities do not mix at linear level in the equations of motion, thus allowing to study them separately also at the level of the action. In the case of scalar perturbations, the background is chosen to satisfy the attractor condition (\ref{eq:attractor}) (with $\alpha>-4$), whereas the analysis of vector and tensor perturbations is valid both on and away from the attractor.

\subsection{Tensor modes}

One may include tensor perturbations by assuming
\begin{align}\
ds_3^2 &= a^2 \left(\delta_{ij} + h_{ij}\right) dx^i dx^j\,,\\
N_i &= 0\,,\\
N &= N(t)\,,
\end{align}
where $\delta^{ij} h_{ij}=0=\delta^{ik}\partial_k h_{ij}$. The Lagrangian quadratic in perturbations then reduces to 
\begin{equation}
\mathcal{L}=\frac{M_\textrm{P}^{2}}{8} \sum_{\epsilon={+,\times}}Na^{3} \left[\frac{\dot{h}_{\epsilon}^{2}}{N^{2}}-\,\frac{1}{a^{2}}\,(\partial_{i}h_{\epsilon})^{2}-\mu_{T}^{2}\,h_{\epsilon}^{2}\right],\label{eq:tenaction}
\end{equation}
where $+$ and $\times$ denote the two different polarizations of tensor perturbations, and with
\begin{equation}
\mu_{T}^2 = \frac{1}{2}\mathcal{X} m^2 \left[r\mathcal{X}\left(\mathcal{X} c_1 + c_2\right)+ \mathcal{X} c_2 + c_3\right].
\end{equation}
The Lagrangian (\ref{eq:tenaction}) is valid both on and away from the attractor (\ref{eq:attractor}). The tensor mass, obtained by looking at the dispersion relation implied by the previous reduced action, is of order of $m$, which is supposed to be of order $H_0$ in order for this theory to explain the late-time acceleration of the universe. We thus have $\mu_T\simeq H_0$. In the following dispersion relation, for the tensor modes,
\begin{equation}
\omega_T^2 = c^2_T\, \frac{k^2}{a^2} + \mu_T^2\,,
\end{equation}
we have $c^2_T = 1$. This means that at all energy scales where $m$ ($\simeq10^{-33}$ eV) is negligible, e.g.\ for astrophysical phenomena, one recovers the same tensor mode phenomenology as in GR. In particular, this theory easily satisfies the bounds imposed by LIGO on the graviton mass, i.e.\ $|\mu_T| < 1.2 \times 10^{-22}$ eV, see e.g.\ \cite{Sakstein:2017xjx}.

\subsection{Vector modes}

The vector perturbations from the metric are given by 
\begin{align}
ds_3^2 &= a^2 \left(\delta_{ij} + \partial_{(i} E_{j)}\right) dx^i dx^j\,,\\
N_i &= B_i\,,\\
N &= N(t)\,,
\end{align}
while among the remaining fields only $\lambda^i$ contributes by
\begin{equation}
\lambda^i = \delta\lambda^i\,,
\end{equation}
since its background value is zero. All vector perturbations are taken to be transverse, i.e.\ $\partial^i E_i = \partial^iB_i = \partial^i\delta\lambda_i = 0$. Using the equations for $\delta\lambda_i$ one may integrate out all degrees of freedom, thus leaving no propagating vector mode. This was expected from the construction of the theory \cite{mqdh3}. Again, this result is valid both on and away from the attractor (\ref{eq:attractor}).

\subsection{Scalar modes}

The scalar perturbations are given in the metric by 
\begin{align}
ds_3^2 &= a^2 (1+ 2\zeta)\delta_{ij}dx^idx^j+ 2 a^2 \partial_i\partial_j s\, dx^i dx^j\,,\\
N_i &= N(t)\partial_i\chi\,,\\
N &= N(t)(1 + \tilde{\alpha})\,,
\end{align}
and in the remaining fields by
\begin{eqnarray}
X & = & X(t)+\delta X\,,\\
\sigma & = & \sigma(t)+\delta\sigma\,,\\
\lambda_{\chi} & = & \lambda_{\chi}(t)+\delta\lambda_{\chi}\,,\\
\lambda_{T} & = & \delta\lambda_{T}\,,\\
\lambda^{i} & = & \frac{1}{a^{2}}\,\partial_{i}\lambda_{V}\,,\\
\lambda & = & \delta\lambda\,.
\end{eqnarray}
Matter fields are also expanded about their background value. In particular, we set $\rho_m=\rho_m^{(0)}(t)+\delta\rho_m$, but drop the superscript $(0)$ from the background quantity for the ease of notation. Here in particular, the background is set to be the attractor solution (\ref{eq:attractor}) which is expected to be reached at early times, as discussed in section \ref{sec:attractor}. After expanding the action quadratically in these linear perturbations, one may integrate out the non-dynamical fields. Thus one obtains an action for propagating fields only, as well as equations setting the auxiliary variables.

\subsubsection{Integration of the non-dynamical fields}

The equation of motion for the field $\lambda_{V}$ sets the following
constraint
\begin{equation}
\zeta=\frac{1}{2}\,\frac{\left[c_{1}\,\left(\alpha+6\right)\mathcal{X}^{2}+2\,c_{2}\,\left(5+\alpha\right)\mathcal{X}+c_{3}\,\left(4+\alpha\right)\right]\delta\sigma}{\Mpl\,\mathcal{X}\,\left(c_{1}\,\mathcal{X}+c_{2}\right)}\,.
\end{equation}

The equation of motion for $\delta\lambda_{\chi}$ sets another constraint,
namely
\begin{equation}
\delta\lambda_{T}=\Mpl\,N\,H\,\tilde{\alpha}-\dot{\delta\sigma}+\frac{N\,\delta X}{\Mpl\,H}\,.
\end{equation}

We will use from later on the gauge-invariant combination
\begin{equation}
\delta\equiv\frac{\delta\rho_m}{\rho_m^{(0)}}+3\,\zeta\,.
\end{equation}

At this point we use the equation of motion for the field $\chi$
to integrate out the field $\tilde{\alpha}$. After replacing $\tilde{\alpha}$,
we integrate out the field $\chi$, via its own equation of motion.
One can use the equation of motion for $\delta\lambda$ in order to
set a constraint for $\delta X$ in terms of the other fields.

After integrating out $\delta X$ and all remaining Lagrange multipliers, the reduced Lagrangian is a function
of $s,\delta\sigma,$ and $\delta$, which can be written as follows
\begin{equation}
\mathcal{L}=A_{11}\,\dot{\delta}^{2}-B_{12}\,(\dot{\delta}\,\delta\sigma-\delta\,\dot{\delta\sigma})-l_{11}\,\delta^{2}-2l_{12}\,\delta\,\delta\sigma-l_{22}\,\delta\sigma^{2}+\beta_{1}\,s^{2}+s\,(\beta_{2}\,\dot{\delta}+\beta_{3}\,\dot{\delta\sigma}+\beta_{4}\,\delta+\beta_{5}\,\delta\sigma)\,.
\end{equation}
One can solve the equation of motion for $s$ as 
\begin{equation}
s=-\frac{1}{2\beta_{1}}\,(\beta_{2}\,\dot{\delta}+\beta_{3}\,\dot{\delta\sigma}+\beta_{4}\,\delta+\beta_{5}\,\delta\sigma)\,,\label{eq:SPaction3fields}
\end{equation}
to integrate out $s$ from the action. 

Then we perform the field redefinition
\begin{eqnarray}
\delta\sigma & = & \delta\sigma_{2}-\frac{\beta_{2}}{\beta_{3}}\,\delta\,,\\
\delta & = & k\,\delta_2\,,
\end{eqnarray}
to find
\begin{eqnarray}
\mathcal{L} & = & k^{2}\,A_{11}\,\dot{\delta}_{2}^{2}-\frac{\beta_{3}^{2}}{4\,\beta_{1}}\,\dot{\delta\sigma}_{2}^{2}+\mathcal{B}\,(\dot{\delta}_{2}\,\delta\sigma_{2}-\delta_{2}\,\dot{\delta\sigma}_{2})-\mathcal{C}_{11}\,\delta{}_{2}^{2}-\mathcal{C}_{22}\,\delta\sigma_{2}^{2}-2\,\mathcal{C}_{12}\,\delta_{2}\,\delta\sigma_{2}\,,\label{eq:SPaction}
\end{eqnarray}
where the order of each coefficient in the high-$k$ limit is 
\begin{equation}
A_{11} = \mathcal{O}(k^{-2})\,,\quad\frac{\beta_{3}^{2}}{4\,\beta_{1}} = \mathcal{O}(k^0)\,,\quad\mathcal{B} = \mathcal{O}(k^{-1})\,,\quad\mathcal{C}_{11} = \mathcal{O}(k^{0})\,,\quad\mathcal{C}_{12} = \mathcal{O}(k^{1})\,,\quad\mathcal{C}_{22} = \mathcal{O}(k^{2})\,.\label{eq:SPaction_terms_order}
\end{equation}
We refer the reader to appendix \ref{sec:appA} for further details in the decomposition of the coefficients $\beta_{n}$ and $l_{mn}$, as well as the terms in (\ref{eq:SPaction_terms_order}).

\subsubsection{Subhorizon limit}

In the high-$k$ limit, we find the following approximated Lagrangian
\begin{equation}
\mathcal{L}\approx\frac{1}{2}\,N\,a^{3}\left[Q_{1}\,\frac{\dot{\delta_2}^{2}}{N^{2}}+Q_{2}\,\frac{\dot{\delta\sigma_2}^{2}}{N^{2}}+\frac{a}{k}\,B\left(\frac{\dot{\delta_2}}{N}\,\delta\sigma_2-\delta_2\,\frac{\dot{\delta\sigma_2}}{N}\right)-L_{11}\,\delta_2^{2}-2\,L_{12}\,\frac{k}{a}\,\delta_2\,\delta\sigma_2-L_{22}\,\frac{k^{2}}{a^{2}}\,\delta\sigma_2^{2}\right],\label{SHaction}
\end{equation}
where $Q_{1}$, $Q_{2}$, $B$, $L_{lm}$ ($l,m\in\{1,2\}$), are all functions of time only. The friction coefficient $B$ is subleading in the quasistatic approximation (see next section).

The kinetic coefficients read
\begin{eqnarray}
Q_{1} &=& a^{2}\,\rho_{m}\,,\\
Q_2 &=& \frac{\Gamma_1^2}{\Gamma_2^2}\frac{q^2}{4Qd^2}\,,
\end{eqnarray}
where we have defined 
\begin{eqnarray}
\Gamma_1 & = & \mathcal{X}\left(c_1\mathcal{X}^2 + 2c_2\mathcal{X}+c_3\right) = \frac{1}{3}\Gamma_{,\mathcal{X}}\mathcal{X}\label{eq:defgamma1}\,,\\
\Gamma_2 & = & \mathcal{X}^2\left(c_1\mathcal{X} + c_2\right) = \frac{1}{6}\Gamma_{,\mathcal{X}\mathcal{X}}\mathcal{X}^2\label{eq:defgamma2}\,,\\
q & = & 2 Q \left[2 H^2 \left(4+\alpha\right)^2 + m^2\Gamma_2\right] - 3m^2 \left[\left(2 - M_\textrm{P}H^2G_{,X}\right)^2\Gamma_2 +(4+\alpha) \left(2 - M_\textrm{P}H^2G_{,X}\right)\Gamma_1\right]\label{eq:defq}, \\
d & = &{m}^{2}\Gamma_{{1}}+2\,{H}^{2} \left( 4+\alpha \right)  \left( M_\textrm{P}H^2G_{,X}-2 \right)\label{eq:defd},\\
Q & = & 3\,\Mpl^{3}\,G_{,XX}H^{4}+\frac{3}{2}\,\Mpl^{2}\,H^{4}\,(G_{,X})^{2}+\Mpl^{2}P_{,XX}H^{2}+6\,\Mpl\,G_{,X}H^{2}+P_{,X}\label{eq:defQ}\,.
\end{eqnarray}
From these, it is possible to read the no ghost conditions in the high-$k$ limit. The no-ghost condition for the field $\delta_2$, reads
\begin{equation}
\rho_{m}>0\,,
\end{equation}
which is always trivially satisfied for canonical matter. The other non-trivial no-ghost condition leads to
\begin{equation}
Q >0\,.\label{eq:noghost}
\end{equation}
This latter condition is, in form, equal to the case of de Sitter, studied in \cite{mqdh3}. 

For the ease of reading, we first decompose the mass coefficients in powers of $m$ as
\begin{eqnarray}
L_{11} &= & -\frac{a^2 \rho_m^2}{2 q^2 M_\textrm{P}^2 \Gamma_1^2 }\left(m^0L_{11,0} + m^2L_{11,2} + m^4L_{11,4} \right),\\
L_{12} &= & -\frac{a \rho_m}{2 d q M_\textrm{P}^3 \Gamma_1 \Gamma_2 }\left(m^0L_{12,0} + m^2L_{12,2} + m^4L_{12,4} \right),\\
L_{22} &= & \frac{1}{2 d^2 M_\textrm{P}^2 \Gamma_2^2}\left(m^0L_{22,0} + m^2L_{22,2} + m^4L_{22,4}\right).
\end{eqnarray}
The coefficients are then given by
\begin{eqnarray}
L_{11,0} &= &16 (\alpha +4)^2 H^2 Q^2 \left[H^2 (\xi_2-6 \Gamma_2 \iota_{-g_1})+\frac{2 \Gamma_2^2 (g-3 g_1) \rho_m }{M_\textrm{P}^2 g}\right],\\
L_{11,2} &= & 8 (\alpha +4) \Gamma_1 Q \left[H^2 (3 g_1 \iota_0 \iota_2+2 \Gamma_2 Q \epsilon)+\frac{6 \Gamma_2^2 Q \rho_m}{M_\textrm{P}^2 g}\right],\\
L_{11,4} &= & 3 \Gamma_1^2 g_1 \left[3 g_1 (8 \Gamma_2 \iota_{-g_1}+\xi_1)+8 \Gamma_2 Q \iota_{-(1+\alpha)}-\frac{18 \Gamma_2^2 g_1 g_2 \rho_m}{H^2 M_\textrm{P}^2 g}\right],\\
L_{12,0} &= & 8 (\alpha +4)^2 H^2 Q \left\{H^2 M_\textrm{P}^2 \Bigl[(g_1+2) \left(\iota_0^2-6 \Gamma_2 \iota_{-g_1}\right)-2 \Gamma_2 P_{,X} \iota_{-g_1}\Bigr]+\frac{2 \Gamma_2 \iota_{-g_1} (3 g_1-g)\rho_m }{g}\right\},\\
L_{12,2} &= & 2(\alpha + 4) \Gamma_1 M_\textrm{P}^2 \biggl(H^2 \Bigl\{3 g_1^2 \iota_{-6} \iota_0+2 g_1 \left[3 (\alpha +4) \Gamma_1 \iota_{-(6+P_{,X})}+2 \Gamma_2 Q \epsilon\right]-2  Q \iota_0 \iota_{-2(3+\alpha)}\Bigr\}\nonumber\\
& & \left. -\frac{6 \Gamma_2 \rho_m (3 g_1 g_2 \iota_0+2 Q \iota_{-g_1})}{M_\textrm{P}^2 g}\right),\\
L_{12,4} &= & \Gamma_1^2 M_\textrm{P}^2 \biggl\{12 \Gamma_2 g_1^2 \iota_{-(3+\alpha)}-3 g_1 \left(\iota_{-(3+\alpha)}^2+\Gamma_2^2 \left(-\alpha ^2-6 \alpha +2 P_{,X}+3\right)\right)+4 \Gamma_2 g \iota_{-(1+\alpha)}\nonumber\\
& &\left.-\frac{18 \Gamma_2^2 g_1 g_2 \rho_m}{H^2 M_\textrm{P}^2 g}\right\},
\end{eqnarray}
\begin{eqnarray}
L_{22,0} &=& 4 (\alpha +4)^2 H^2 \biggl(g_1 H^2 M_\textrm{P}^2 \Bigl\{2 \Gamma_2 P_{,X} (\iota_{-g_1}+\iota_0)-(g_1+2) \left[\iota_0^2-6 \Gamma_2 (\iota_{-g_1}+\iota_0)\right]\Bigr\}\nonumber\\
& &\left.+2 \rho_m \left[\frac{3}{g} (g_1\iota_{-g_1}^2 + g_2 \iota_0^2)-\iota_{-g_1}^2 \right]\right),\\
L_{22,2} &=& 4 (\alpha +4) \Gamma_1 \biggl(H^2 M_\textrm{P}^2 \Bigl\{g_1 (\iota_{2-2\alpha} \iota_{-g_1}-\Gamma_2 g_1 \iota_{-5})-(\alpha +4) \Gamma_1 \left[\iota_0-2 \Gamma_2 (P_{,X}+6)\right]\Bigr\}\nonumber\\
& &\left.+\frac{\rho_m}{g} \left[6 (\alpha +4) \Gamma_1 \Gamma_2 g_2-3 \iota_{-g_1}^2\right]\right),\\
L_{22,4} &=& \frac{\Gamma_1^2}{H^2} \left(H^2 M_\textrm{P}^2 \xi_3+6 \Gamma_2^2 g_2 \frac{\rho_m}{g}\right),
\end{eqnarray}
where we have defined $\iota_n \equiv (4+\alpha) \Gamma_1 + n \Gamma_2 $, for $n\in \mathbb{R}$, and
\begin{eqnarray}
g_1 & = & G_{,X}H^2M_\textrm{P}-2\,,\\
g_2 & = & 2 G_{,X}H^2M_\textrm{P} + G_{,XX}H^4M_\textrm{P}^3\,,\\
\epsilon &=& 2 (\alpha+4) \Gamma_1 - (3+2 \alpha)\Gamma_2\\
\xi_1 & = &(\alpha +4)^2 \Gamma_1^2 - 4(\alpha +4)\Gamma_1 \Gamma_2 -2(P_{,X}+6)\Gamma_2^2 \\
\xi_2 & = &(\alpha +4)^2 \Gamma_1^2 + 6(\alpha +4)\Gamma_1 \Gamma_2 +2(P_{,X}+6)\Gamma_2^2 \\
\xi_3 & = &3 (\alpha +4)^2 \Gamma_1^2-4 (\alpha +4)(\alpha +g_1+2)\Gamma_1 \Gamma_2 +2 (2 (\alpha +3) g_1+P_{,X}+6)\Gamma_2^2 \\
g & = & P_{,X} - 6 + 12 G_{,X}H^2 M_\textrm{P} + 3 G_{,XX}H^4M^3_\textrm{P} + H^2 M^2_\textrm{P}P_{,XX} = Q-\frac{3}{2}g_1^2\,,
\end{eqnarray}
and $\Gamma_1$, $\Gamma_2$, $q$, $d$, and $Q$ are given in equations (\ref{eq:defgamma1})-(\ref{eq:defQ}). From the mass coefficients, it is possible to read the scalar sound speed in the high-$k$ limit. In particular, for modes for which $\omega^{2}=c_{s}^{2}\,\frac{k^{2}}{a^{2}}$,
we have
\begin{equation}
\left(\omega^{2}\,Q_{2}-L_{22}\,\frac{k^{2}}{a^{2}}\right)\delta\sigma_2\approx0\,,
\end{equation}
so that
\begin{equation}
c_{s}^{2}=\frac{L_{22}}{Q_{2}}\,, \label{eqn:cs2}
\end{equation}
whereas dust still has zero speed of propagation.

\subsubsection{Quasistatic approximation}

Using the action (\ref{SHaction}), one obtains the following equations of motion
\begin{eqnarray}
&&-\frac{d}{dt}\!\left(a^{3}\,Q_{1}\,\frac{{{\dot\delta}_2}}{N}\right)-\frac{1}{2}\,\frac{d}{dt}\!\left(\frac{a^{4}}{k}\,B\,\delta\sigma_2\right)-\frac{a^{4}}{2k}\,B\dot{\delta\sigma_2}-N\,a^{3}\,L_{11}\,\delta_2-N\,a^{3}\,L_{12}\,\frac{k}{a}\,\delta\sigma_2 = 0\,,\\
&&-\frac{d}{dt}\left(a^{3}\,Q_{2}\,\frac{\dot{\delta\sigma_2}}{N}\right)+\frac{a^{4}}{2k}\,B\,{{\dot\delta}_2}+\frac{1}{2}\,\frac{d}{dt}\!\left(\frac{a^{4}}{k}\,B\,\delta_2\right)-N\,a^{3}\,L_{12}\,\frac{k}{a}\,\delta_2-N\,a^{3}\,L_{22}\,\frac{k^{2}}{a^{2}}\,\delta\sigma_2 = 0\,.
\end{eqnarray}
In the quasistatic approximation, where $\frac{\ddot{\delta\sigma_2}}{N^{2}}\simeq H\,\frac{\dot{\delta\sigma_2}}{N}\simeq H^{2}\,\delta\sigma_2\ll\frac{k^{2}}{a^{2}}\,\delta\sigma_2$, these two equations become 
\begin{gather}
L_{12}\,\delta_2+L_{22}\,\frac{k}{a}\,\delta\sigma_2 \approx 0\,, \label{eqn:sigma2-quasistatic}\\
\frac{1}{N\,a^{3}}\,\frac{d}{dt}\left(a^{3}\,Q_{1}\,\frac{{{\dot\delta}_2}}{N}\right)+L_{11}\,\delta_2+L_{12}\,\frac{k}{a}\,\delta\sigma_2 \approx 0\,. \label{eqn:delta2-quasistatic}
\end{gather}
By solving (\ref{eqn:sigma2-quasistatic}) for $\delta\sigma_2$ and substituting it to (\ref{eqn:delta2-quasistatic}), one obtains the equation for $\delta_2$ as
\begin{equation}
\frac{1}{N}\,\frac{d}{dt}\left(\frac{\dot{\delta_2}}{N}\right)+2H\,\frac{\dot{\delta_2}}{N}+\frac{1}{\rho_{m}a^{2}}\left(L_{11}-\frac{L_{12}^{2}}{L_{22}}\right)\delta_2\approx0\,.
\end{equation}
By comparing this with the definition of $G_\textrm{eff}$,
\begin{equation}
\frac{1}{N}\frac{d}{dt}\left(\frac{\dot{\delta}}{N}\right) + 2 H \frac{\dot{\delta}}{N} - 4\pi G_\textrm{eff} \rho_m \delta = 0\,,
\end{equation}
one can obtain
\begin{equation}
\frac{G_\textrm{eff}}{G_N} = \frac{2M_\textrm{P}^2}{\rho_m^2a^2}
\left(\frac{L^2_{12}}{L_{22}}-L_{11}\right). \label{eq:GeffoverGN}
\end{equation}
This ratio is generally different from one.

In this same quasistatic approximation it is possible to show that the two gauge invariant Bardeen potentials
\begin{eqnarray}
  \Psi&=&\tilde\alpha+\frac{\dot\chi}N-\frac1N\,\frac{d}{dt}\!\left(\frac{a^2\dot s}{N}\right),\\
  \Phi&=&-\zeta-H\,\chi+a^2\,H\,\frac{\dot s}N\,,
\end{eqnarray}
both satisfy the Poisson equations
\begin{eqnarray}
  -\frac{k^2}{a^2}\,\Psi&=&\frac1{2\Mpl^2}\,\frac{G_{\rm eff}}{G_N}\,\rho_m\,\delta\,,\\
  -\frac{k^2}{a^2}\,\Phi&=&\frac1{2\Mpl^2}\,\frac{G_{\Phi}}{G_N}\,\rho_m\,\delta\,,
\end{eqnarray}
where the ratio $G_\textrm{eff}/G_N$ is given in Eq. (\ref{eq:GeffoverGN}), whereas 
\begin{equation}
\frac{G_{\Phi}}{G_{N}} = \frac{g_{\Phi,n,4}m^4+g_{\Phi,n,2}m^2+g_{\Phi,n,0}}{g_{\Phi,d,4}m^4+g_{\Phi,d,2}m^2+g_{\Phi,d,0}}
\end{equation}
with the following coefficients
\begin{eqnarray}
g_{\Phi,n,4} & = & -2 \Gamma_1^2 H^2 \iota_2 M_\textrm{P}^2 g \iota_{-(1+\alpha)}\,,\\
g_{\Phi,n,2} & = & -4 (\alpha +4) \Gamma_1 H^2 \left(3 \rho_m \left[\Gamma_2 g_1 \iota_2+\Gamma_2 (g_2-2) \iota_0-\iota_0^2\right]\right.\nonumber\\
& & \left.+H^2 M_\textrm{P}^2 g \left\{g_1 \left[\iota_0 \iota_{6-\alpha}-2 (2 \alpha +3) \Gamma_2^2\right]+\iota_0 \left[\Gamma_2 (2 \alpha +P_{,X}+12)-\iota_0\right]\right\}\right),\\
g_{\Phi,n,0} & = & -8 (\alpha +4)^2 H^4 \left\{\rho_m \left[\iota_2 \iota_{-g_1} (3 g_1-g)+3 g_2 \iota_0^2\right]\right.\nonumber\\
& &\left. +H^2 M_\textrm{P}^2 g \left[\Gamma_2 g_1 \iota_2 (3 g_1+P_{,X}+6)+\iota_0 (g_1 \iota_{-6}+2 \iota_{-(6+P_{,X})})\right]\right\},\\
g_{\Phi,d,4} & = & -\Gamma_1^2 \left[H^2 M_\textrm{P}^2 g (4 \iota_{-(3+\alpha)} \iota_{-g_1}-\xi_1)+6 \Gamma_2^2 g_2 \rho_m\right],\\
g_{\Phi,d,2} & = & 12 (\alpha +4) \Gamma_1 H^2 \rho_m \left[\iota_{-g_1}^2-2 (\alpha +4) \Gamma_1 \Gamma_2 g_2\right]\nonumber\\
&& -4 (\alpha +4) \Gamma_1 H^4 M_\textrm{P}^2 g \left\{(g_1+2) \left[-2 \alpha  \Gamma_2 \iota_{-g_1}+3 \Gamma_2 (\iota_{g_1}+\iota_2)+\iota_0 \iota_{-2(2+g_1)}\right]-\xi_3 +2 \Gamma_2 \iota_1 P_{,X}\right\},\\
g_{\Phi,d,0} & = & -24 (\alpha +4)^2 H^4 \rho_m \left[g_1 \iota_{-g_1}^2+(\alpha +4)^2 \Gamma_1^2 g_2\right]+8 (\alpha +4)^2 H^4 \rho_m g \iota_{-g_1}^2\nonumber\\
&&+4 (\alpha +4)^2 g_1 H^6 M_\textrm{P}^2 g \left\{(g_1+2) \left[\iota_0^2-6 \Gamma_2 (\iota_{-g_1}+\iota_0)\right]-2 \Gamma_2 P_{,X} (\iota_{-g_1}+\iota_0)\right\}.
\end{eqnarray}
By combining both equations, one directly finds the slip parameter $\eta$, simply given by
\begin{equation}
\eta \equiv \frac{\Psi}{\Phi} = \frac{G_\textrm{eff}}{G_\Phi}\,.\label{eq:eta}
\end{equation}

Although we have given the general expressions for $G_{\rm eff}$ and $\eta$, and their values do depend on the choice of the model and the background, we can study, for example, their behavior at early times, i.e.\ when $\rho_m\to3\Mpl^2\,H^2$ and $m^2/H^2\to0$, in the case $|\Gamma_2|\ll 1$, $G_{,X}H^2 M_\textrm{P}\to0$, and $G_{,XX}H^4M^3_\textrm{P}\to0$. In fact, in this case we find
\begin{eqnarray}
  G_{\rm eff}&\to& G_N\,,\\
  \eta&\to& 1\,,\\
  Q_2&\to&\frac { \left( \alpha+4 \right) ^{2} \left( \Mpl^2{H}^{2}P_{{{,XX}}}+P_{{,X}} \right) \Gamma_1^2}%
           {4\Gamma_2^2}\,,\\
  c_s^2&\to&\frac{3}{6-\Mpl^{2}{H}^{2}P_{{{, XX}}}-P_{{,X}}}\, .
\end{eqnarray}
This results shows that there is parameter space for which it is possible to have at early times a GR limit in the behavior of the growth of structure, together with a stable background, provided that $0<\Mpl^2{H}^{2}P_{{{,XX}}}+P_{{,X}}<6$.

\section{Summary and discussion}\label{sec:discussion}

In \cite{mqdh3}, it was shown that the minimal theory of quasidilaton theory could be extended with a Horndeski structure to accommodate the Vainshtein screening, while retaining important features at the cosmological level. Among these are the existence of FLRW cosmologies and in particular the presence of a late-time de Sitter attractor, generally stable under perturbations.
Here, we extend that work by showing that the dynamics of the quasidilaton scalar allow for an attractor that generalizes the attractor solution to the system with matter. This was expected since the attractor condition depends on the equation of motion for $\lambda$, which does not couple directly to the matter sector. The attractor thus obtained is shown to be stable in a wide range of parameters under inhomogeneous perturbations. It is sufficient to satisfy the no-ghost condition (\ref{eq:noghost}) and the positivity of the squared sound speed (\ref{eqn:cs2}), in order to avoid ghost- and gradient-instabilities. We have then found that the speed limit of gravitational waves coincides with the speed of light for any homogeneous and isotropic cosmological background, on or away from the attractor. Furthermore, we have studied scalar perturbations of this theory in the quasistatic approximation and found a modified gravitational constant $G_{\rm eff}$ and slip parameter $\eta$ for matter perturbations as shown in Eqs.\ (\ref{eq:GeffoverGN}) and (\ref{eq:eta}). 

This last result might be of particular interest as it is well known that it is in general hard to have a non-trivial modified-gravitational law (especially when we look for theories for which $G_{\rm eff}<G_N$), which could be in principle able to explain weak gravity. It was shown in \cite{DeFelice:2016ufg}, that weak gravity can be easily achieved in the normal branch of MTMG. The minimal theory of quasidilaton massive gravity with cubic Horndeski Lagrangian, while keeping the speed of the tensor modes equal to unity, might still then be able to achieve weak gravity. Although more work is needed in order to study this interesting possibility, the phenomenology of this theory looks promising in the light of future experiments which can shed light on the behavior of gravity, e.g.\ in redshift-distortion-data.

In future work, it will be interesting to make contact with observations at shorter scales. While finding the modified gravitational constant was a first step in this regard, several other parameters are relevant to observations. These should in principle be obtainable in the context of the minimal theory of quasidilaton massive gravity. Due to the presence of the quasidilaton field, one originally \cite{mqd} expected the presence of a fifth force, thus potentially compromising the agreement of the theory at e.g.\ Solar System scales. Thanks to the extension by Horndeski-type terms \cite{mqdh3} this can in principle be avoided by the Vainshtein effect. It remains important to study how this Vainshtein mechanism works to screen the potential fifth force in detail.

\acknowledgments
We express special thanks to Fran\c{c}ois Larrouturou for many valuable discussions and comments, technical assistance, as well as proofreading of the present work. ADF was supported by JSPS KAKENHI Grant Number 16K05348. The work of SM was supported by Japan Society for the Promotion of Science (JSPS) Grants-in-Aid for Scientific Research (KAKENHI) No. 17H02890, No. 17H06359, and by World Premier International Research Center Initiative (WPI), MEXT, Japan.  MO acknowledges the support from the Japanese Government (MEXT) Scholarship for Research Students.

\appendix

\section{Coefficients in the scalar perturbations Lagrangian}\label{sec:appA}

The coefficients in the Lagrangian (\ref{eq:SPaction}) can be expressed in terms of the coefficients $\beta_{n}$ and $l_{mn}$, defined in (\ref{eq:SPaction3fields}), as 
\begin{eqnarray}
\mathcal{B} & = & \frac{\left[-4\,B_{12}\beta_{1}+\left(-\beta_{5}+\dot{\beta}_{3}\right)\beta_{2}+\beta_{3}\left(\beta_{4}-\dot{\beta}_{2}\right)\right]k}{4\beta_{1}}\,,\\
\mathcal{C}_{22} & = & l_{{22}}-\frac{\beta_{{3}}\dot{\beta}_{5}}{4\beta_{{1}}}+\frac{\beta_{{5}}^{2}}{4\beta_{{1}}}-\frac{\dot{\beta}_{3}\,\beta_{{5}}}{4\beta_{{1}}}+\frac{\beta_{{3}}\beta_{{5}}\dot{\beta}_{1}}{4\beta_{{1}}^{2}}\,,\\
\mathcal{C}_{11} & = & \frac{{k}^{2}\dot{\beta}_{3}\beta_{{2}}\beta_{{4}}}{2\beta_{{1}}\beta_{{3}}}+{k}^{2}l_{{11}}+\frac{{k}^{2}l_{{22}}{\beta_{{2}}^{2}}}{{\beta_{{3}}}^{2}}-\frac{{k}^{2}l_{{12}}\beta_{{2}}}{2\beta_{{3}}}+\frac{{k}^{2}{\beta_{{2}}^{2}}{\beta_{{5}}^{2}}}{4\beta_{{1}}{\beta_{{3}}}^{2}}-\frac{{k}^{2}\beta_{{2}}\beta_{{4}}\beta_{{5}}}{2\beta_{{1}}\beta_{{3}}}+\frac{{k}^{2}\beta_{{4}}^{2}}{4\beta_{{1}}}\nonumber \\
 &  & {}+\frac{{k}^{2}\dot{\beta}_{2}^{2}}{4\beta_{{1}}}-\frac{{k}^{2}\dot{\beta}_{2}\dot{\beta}_{3}\beta_{{2}}}{2\beta_{{1}}\beta_{{3}}}+\frac{k^{2}\dot{\beta}_{2}B_{{12}}}{\beta_{{3}}}+\frac{{k}^{2}\dot{\beta}_{2}\beta_{{2}}\beta_{{5}}}{2\beta_{{1}}\beta_{{3}}}-\frac{k^{2}\dot{\beta}_{2}\beta_{{4}}}{2\beta_{{1}}}+\frac{{k}^{2}\dot{\beta}_{3}^{2}\beta_{{2}}^{2}}{4\beta_{{1}}{\beta_{{3}}}^{2}}-\frac{{k}^{2}\dot{\beta}_{3}B_{{12}}\beta_{{2}}}{{\beta_{{3}}}^{2}}-\frac{{k}^{2}\dot{\beta}_{3}\beta_{{2}}^{2}\beta_{{5}}}{2\beta_{{1}}{\beta_{{3}}}^{2}}\,,\\
\mathcal{C}_{12} & = & kl_{{12}}-\frac{kl_{{22}}\beta_{{2}}}{\beta_{{3}}}-\frac{k\beta_{{2}}\ddot{\beta}_{3}}{8\beta_{{1}}}-\frac{k\beta_{{2}}\beta_{{5}}^{2}}{4\beta_{{3}}\beta_{{1}}}+\frac{k\beta_{{3}}\ddot{\beta}_{2}}{8\beta_{{1}}}+\frac{k\beta_{{4}}\beta_{{5}}}{4\beta_{{1}}}\nonumber \\
 &  & {}-\frac{k\beta_{{3}}\dot{\beta}_{1}\dot{\beta}_{2}}{8{\beta_{{1}}}^{2}}+\frac{k\dot{\beta}_{1}\dot{\beta}_{3}\beta_{{2}}}{8{\beta_{{1}}}^{2}}-\frac{k\dot{\beta}_{1}\beta_{{2}}\beta_{{5}}}{8{\beta_{{1}}}^{2}}+\frac{k\beta_{{3}}\dot{\beta}_{1}\beta_{{4}}}{8{\beta_{{1}}}^{2}}-\frac{k\dot{\beta}_{2}\beta_{{5}}}{8\beta_{{1}}}-\frac{k\beta_{{3}}\dot{\beta}_{4}}{8\beta_{{1}}}+\frac{k\dot{\beta}_{3}\beta_{{2}}\beta_{{5}}}{4\beta_{{3}}\beta_{{1}}}-\frac{k\dot{\beta}_{3}\beta_{{4}}}{8\beta_{{1}}}+\frac{k\dot{\beta}_{5}\beta_{2}}{8\beta_{{1}}}\,.
\end{eqnarray}

One may further study the decomposition of $\beta_{n}$ and $l_{mn}$ in powers of $k$,
\begin{eqnarray}
\beta_{1} & = & \frac{\beta_{11}\,k^{6}+\beta_{12}\,k^{4}}{\beta_{13}k^{2}+\beta_{14}}\,,\\
\beta_{2} & = & \frac{\beta_{21}\,k^{2}}{\beta_{22}\,k^{2}+\beta_{23}}\,,\\
\beta_{3} & = & \frac{\beta_{31}\,k^{2}}{\beta_{32}}\,,\\
\beta_{4} & = & \frac{\beta_{41}k^{4}+\beta_{42}k^{2}}{\beta_{43}k^{2}+\beta_{44}}\,,\\
\beta_{5} & = & \frac{\beta_{51}\,k^{6}+\beta_{52}\,k^{4}+\beta_{53}\,k^{2}}{\beta_{54}k^{2}+\beta_{55}}\,,\\
l_{11} & = & \frac{l_{111}k^{4}+l_{112}k^{2}+l_{113}}{l_{114}\,k^{4}+l_{115}\,k^{2}+l_{116}}\,,\\
l_{12} & = & \frac{l_{121}k^{6}+l_{122}k^{4}+l_{123}\,k^{2}+l_{124}}{l_{125}\,k^{4}+l_{126}\,k^{2}+l_{127}}\,,\\
l_{22} & = & \frac{l_{221}k^{6}+l_{222}k^{4}+l_{223}\,k^{2}+l_{224}}{l_{225}\,k^{2}+l_{226}}\,,\\
B_{12} & = & \frac{B_{121}\,k^{2}+B_{122}}{B_{123}\,k^{2}+B_{124}}\,,
\end{eqnarray}
where the objects $\beta_{mn}$, $l_{mno}$, and $B_{mno}$ are functions of time only. 

Let us now calculate $\mathcal{B}$. We find that at highest
order in $k$
\[
\mathcal{B}=\frac{k\left(\beta_{{13}}\left(-\beta_{{43}}\beta_{{51}}\beta_{{21}}\beta_{{32}}+\beta_{{41}}\beta_{{54}}\beta_{{22}}\beta_{{31}}\right)B_{{123}}-4\,\beta_{{32}}\beta_{{11}}\beta_{{22}}\beta_{{54}}\beta_{{43}}B_{{121}}\right)}{4\beta_{{32}}\beta_{{22}}\beta_{{11}}\beta_{{54}}\beta_{{43}}B_{{123}}}=0\cdot k + \mathcal{O}(k^{-1})\,,
\]
so that we need to look for the next leading order, which is of order
$\mathcal{O}(1/k)$, that is $\mathcal{B}=a^4B(t)/(2k)+\mathcal{O}(k^{-3})$.

Let us now calculate $\mathcal{C}_{22}$. We find
\[
\mathcal{C}_{22}=\frac{k^{4}\left(4\,l_{{221}}\beta_{{11}}\beta_{{54}}^{2}+l_{{225}}\beta_{{13}}\beta_{{51}}^{2}\right)}{4\beta_{{11}}l_{{225}}\beta_{{54}}^{2}}=0\cdot k^4+\mathcal{O}(k^2)\,,
\]
on substituting the equations of motion. Then the next term is of
order $\mathcal{O}(k^{2})$, as in
\begin{eqnarray}
\mathcal{C}_{22} & = & \frac{k^{2}}{2\beta_{{11}}^{2}l_{{225}}^{2}\beta_{{54}}^{3}\beta_{{32}}^{2}}\Bigl[l_{225}^{2}\Bigl\{\beta_{51}\left(\left[\left(\beta_{52}\beta_{13}+\frac{1}{2}\,\beta_{51}\beta_{14}\right)\beta_{11}-\frac{1}{2}\,\beta_{12}\beta_{13}\beta_{51}\right]\beta_{54}-\beta_{11}\beta_{13}\beta_{55}\beta_{51}\right)\beta_{32}^{2}\nonumber \\
 &  & {}+\frac{1}{2}\,\left(\left\{ \left[\left(-\dot{\beta}_{13}\beta_{31}-\dot{\beta}_{31}\beta_{13}\right)\beta_{51}-\dot{\beta}_{51}\beta_{13}\beta_{31}\right]\beta_{11}+\beta_{51}\dot{\beta}_{11}\beta_{13}\beta_{31}\right\} \beta_{54}+\beta_{51}\dot{\beta}_{54}\beta_{11}\beta_{13}\beta_{31}\right)\beta_{54}\beta_{32}\nonumber \\
 &  & {}+\frac{1}{2}\,\beta_{{31}}\beta_{{11}}\beta_{{13}}\beta_{{54}}^{2}\beta_{{51}}\dot{\beta}_{32}\Bigr\}+2\,l_{{225}}l_{{222}}\beta_{{32}}^{2}\beta_{{11}}^{2}\beta_{{54}}^{3}-2\,l_{{226}}l_{{221}}\beta_{{32}}^{2}\beta_{{11}}^{2}\beta_{{54}}^{3}\Bigr]\,.
\end{eqnarray}

Along the same lines we find that the leading order term for $\mathcal{C}_{11}$ 
which is of order $k^{2}$ vanishes, so that we need to evaluate the
next leading order which is of order $\mathcal{O}(k^{0})$. As for
the term $\mathcal{C}_{12}$, we have that the leading order term,
which is of order $k^{3}$, vanishes and thus we need to evaluate the
next leading order which is of order $\mathcal{O}(k)$.

\end{document}